# HIGH-GRADIENT SRF CAVITY WITH MINIMIZED SURFACE E.M. FIELDS AND SUPERIOR BANDWIDTH FOR THE ILC

N. Juntong and R.M. Jones
The Cockcroft Institute, Daresbury, Warrington, Cheshire WA4 4AD, UK.
The University of Manchester, Oxford Road, Manchester, M13 9PL, UK.

*Abstract*

Results are presented on an alternative cavity to the ILC baseline design of TESLA-style SRF main accelerating linacs. This re-optimised shape enhances the bandwidth of the accelerating mode and has reduced surface electric and magnetic fields, compared to the baseline design and some current high gradient designs. Detailed simulations on the e.m. fields for the New Low Surface Field (NLSF) cavity, including end-cell and coupler designs, are reported. The re-distributed dipole modes are also discussed.

## INTRODUCTION

It is anticipated that the successor to the LHC [1-2] will be a linear collider to accelerate electrons and positrons. The ILC [3] design aims at colliding leptons at an initial center of mass energy of 500 GeV with a proposed later upgrade to 1 TeV. The superconducting cavities in the main accelerating linacs of the ILC are based on the TESLA [4] design. The baseline design aims at an average accelerating gradient of 31.5 MV/m. However, other designs exist with the potential for higher accelerating gradients. Increasing the accelerating gradient is desirable, as it raises the overall efficiency of the machine. Re-entrant (RE) [5], Low-loss (LL) [6] and Ichiro (IR) [7] are candidates for higher gradient cavities. These designs aim at producing accelerating gradients of ~50 MV/m within 9-cell cavities. Single cells have achieved gradients in excess of 50 MV/m. Indeed at Cornell, a RE cell achieved 52 MV/m [8] and Low Loss (LL) cells at KEK have obtained between 45 to 51 MV/m [9]. These designs are focussed on minimising the ratio of the surface e.m. fields to the accelerating gradient. In particular, the ratio of the surface magnetic field to accelerating gradient ($B_s/E_a$) has been minimised by suitably shaping the walls of the cavity. The critical surface magnetic field is in the range 180 -230mT [10]. Another recent design incorporates minimising an additional quantity, the ratio of the surface electric field to surface accelerating field ($E_s/E_a$) and this is the Low Surface Field design (LSF) [11]. However, the bandwidth of the accelerating mode in LSF design is reduced by ~18% compared to the LL cavity. This reduces the overall stability of the cavity as the frequency separation of modes is proportional to the bandwidth [12].

In order to investigate means to enhance the bandwidth a design study has been pursued which focuses on optimising three quantities: minimisation of $E_s/E_a$ and $B_s/E_a$ whilst maximising $k_c$. This has resulted in a new design, the New Low Surface Field (NLSF) [12] cavity, based on LL and LSF geometries. Detailed simulations on the e.m. fields for the middle cells of NLSF were reported in [12]. The NLSF shape has comparable surface e.m. fields ratio to that in the LSF cavity, but with an enhanced fractional bandwidth. Here we report on the modes and e.m. fields in a complete 9-cell NLSF cavity (including end-cells and couplers) and on additional optimised designs based on re-entrant and TESLA shapes.

This paper is organized such that the optimization strategy and results are described in the next section, followed by detailed simulations on the field flatness of the accelerating mode. The penultimate section concerns the properties of the higher order dipole modes and an initial design of the fundamental mode coupler. The final main section includes some concluding remarks.

## OPTIMISATION STRATEGY

The parameters used to optimise $E_s/E_a$, $B_s/E_a$ and $k_c$ for a single cell are illustrated in Fig. 1. The cell consists of two elliptical surfaces, iris and equator, connected by a surface which is at angle of θ with respect to the vertical axis. The ellipticity of each surface is determined by the ratio of *a/b* and *A/B*. For the current NLSF [12] cavity, these two surfaces are connected with a vertical surface (θ=0). A Mathematica [13] code was written to run the Superfish [14] code to facilitate an investigation of the surface fields as the geometrical parameters are varied.

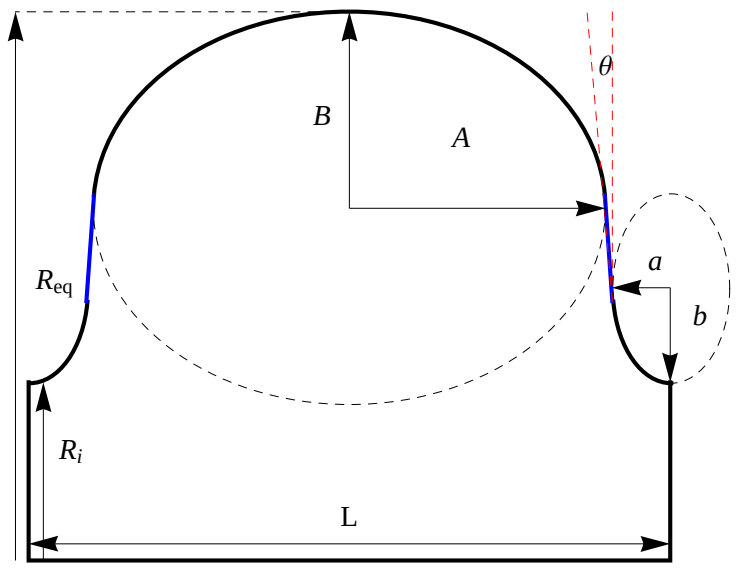

Figure 1: Single cell parameterisation.

Superfish was operated in the field solver mode to seek the 0 and π phase advance per cell frequencies. In this manner the fractional bandwidth and surface field are obtained. In all simulations, the mesh spacing was maintained at ~0.02λ (λ being the operating wavelength). Results were obtained in which the iris elliptical parameter *b* is varied, whilst *a* is maintained at a fixed value. The equator elliptical parameter *B* is used to ensure the cell is tuned to the accelerating frequency of 1.3 GHz to within a tolerance of 100 kHz. Contours plots of three figures of merit: $E_s/E_a$, $B_s/E_a$ and $k_c$, for a NLSF-like shape are displayed in Figs. 2-4. For the sake of completeness, the original NLSF [12] shape is indicated by red points.

It is clear from these results that the surface magnetic field $B_s$ and the fractional bandwidth $k_c$ both decrease as the iris thickness 2*a* is decreased. As expected, the surface electric field $E_s$ behaves in the opposite manner.

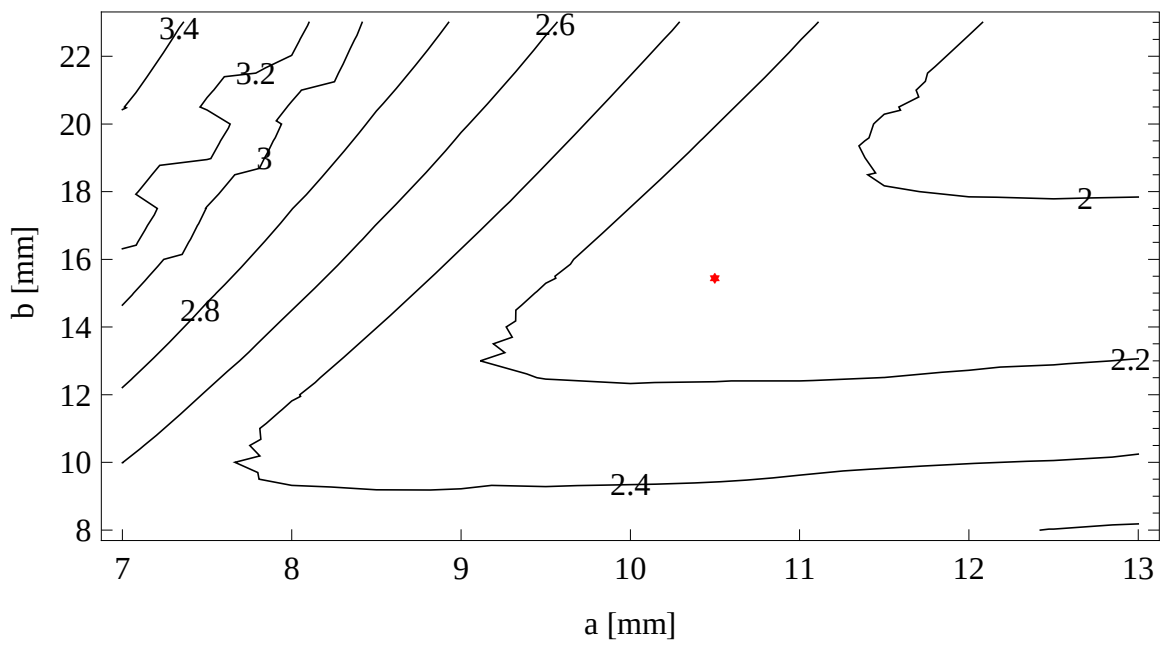

Figure 2: Contours of $E_s/E_a$ as a function of the iris parameters a and b for the NLSF-like shape. The iris radius is maintained at a fixed value of 32 mm.

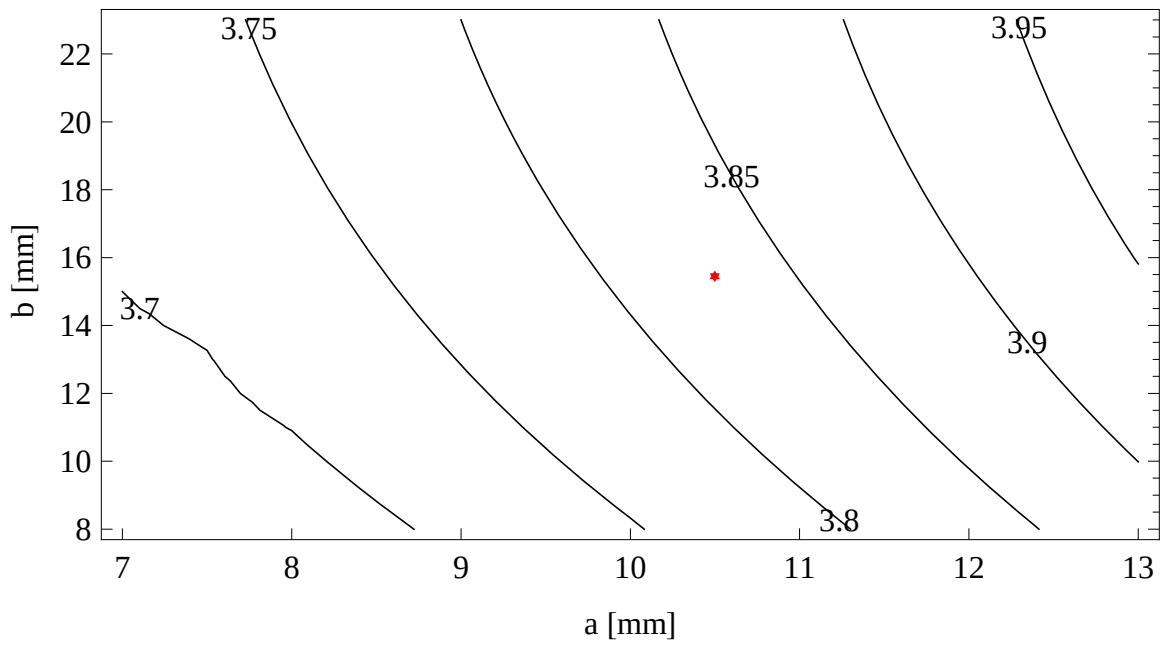

Figure 3: Contours of $B_s/E_a$ for parameters given in Fig. 2.

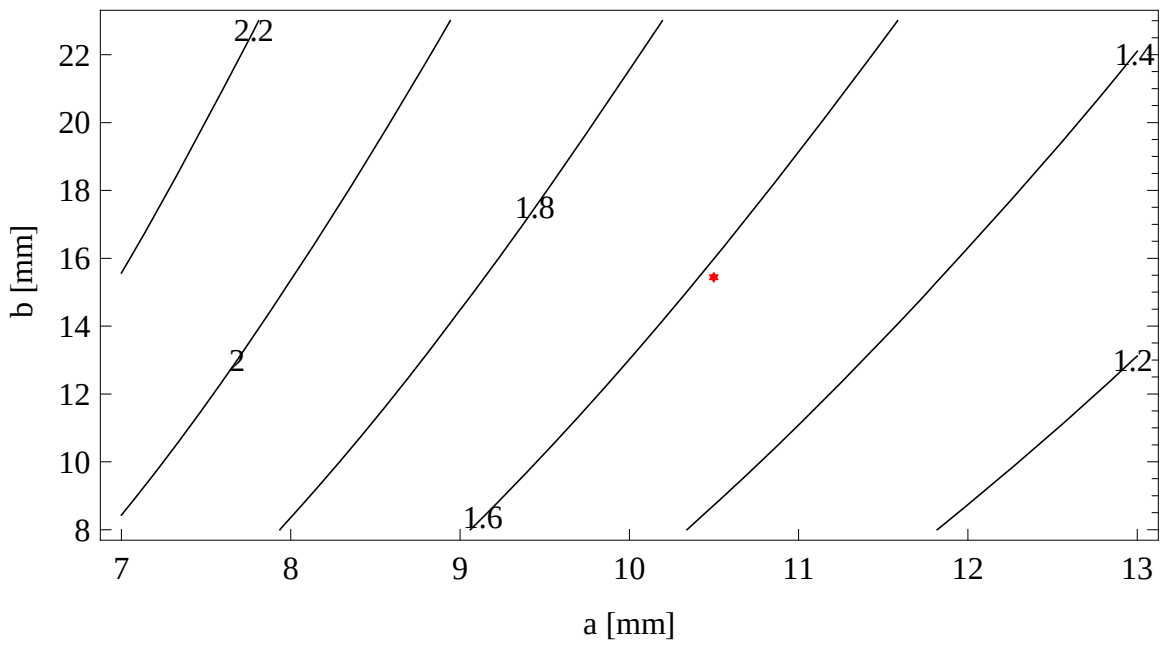

Figure 4: Bandwidth $k_c$ for parameters given in Fig. 2.

We performed additional simulations to investigate the influence of varying the iris radius on the three figures of merit. To this end the iris radius was varied in the range 30 to 33 mm. The results of these simulations are summarised in Fig 5, along with those from existing cavities. It is evident that the bandwidth $k_c$ is very sensitive to changes in the iris radius. The surface fields are however quite unaffected by iris variations in this range.

Selected designs have been validated with HFSS v11 [15]. The discrepancy between Superfish and HFSS results for $k_c$, $E_s/E_a$ and $B_s/E_a$ are 1.7%, 8.2% and 1.3 % respectively. One particular design, NLSF-A, with a 31 mm iris radius and 10 mm iris thickness, has a lower $B_s/E_a$ ratio than the NLSF shape and a comparable ratio of $E_s/E_a$ ratio, but this shape has a reduced value of $k_c$. Table 1 summarizes the final designs

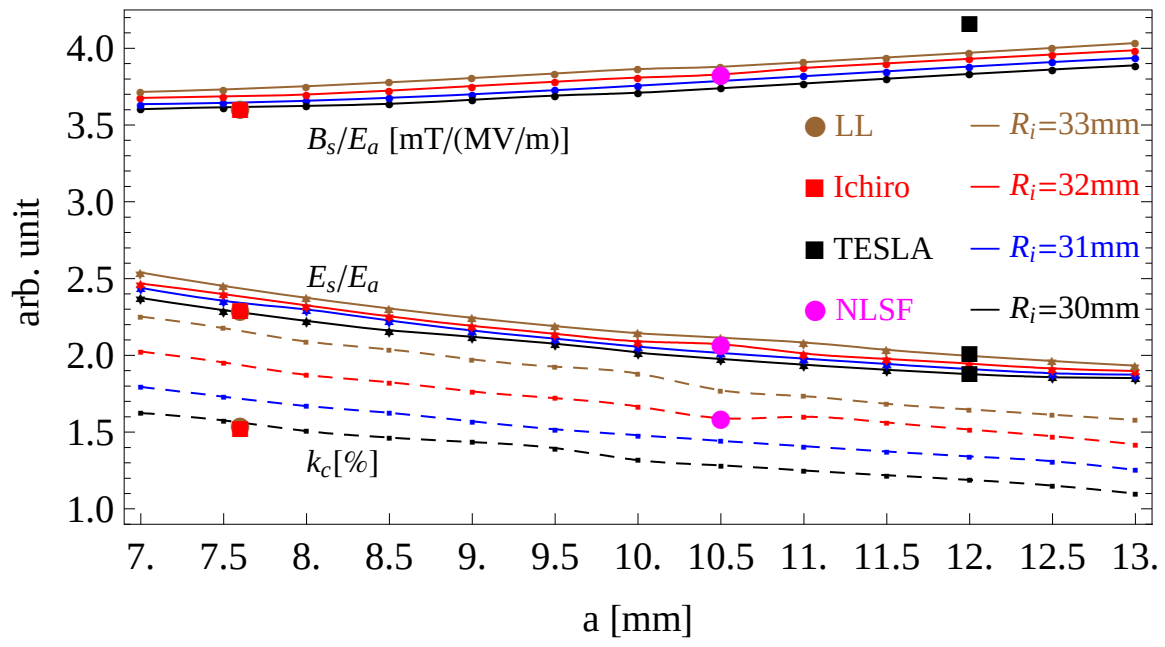

Figure 5: All three figures of merit versus variation in iris thickness for several iris radii. Solid lines indicate the surface e.m. field, which has been minimized, whereas dashed lines refer to the bandwidth, which has been maximized.

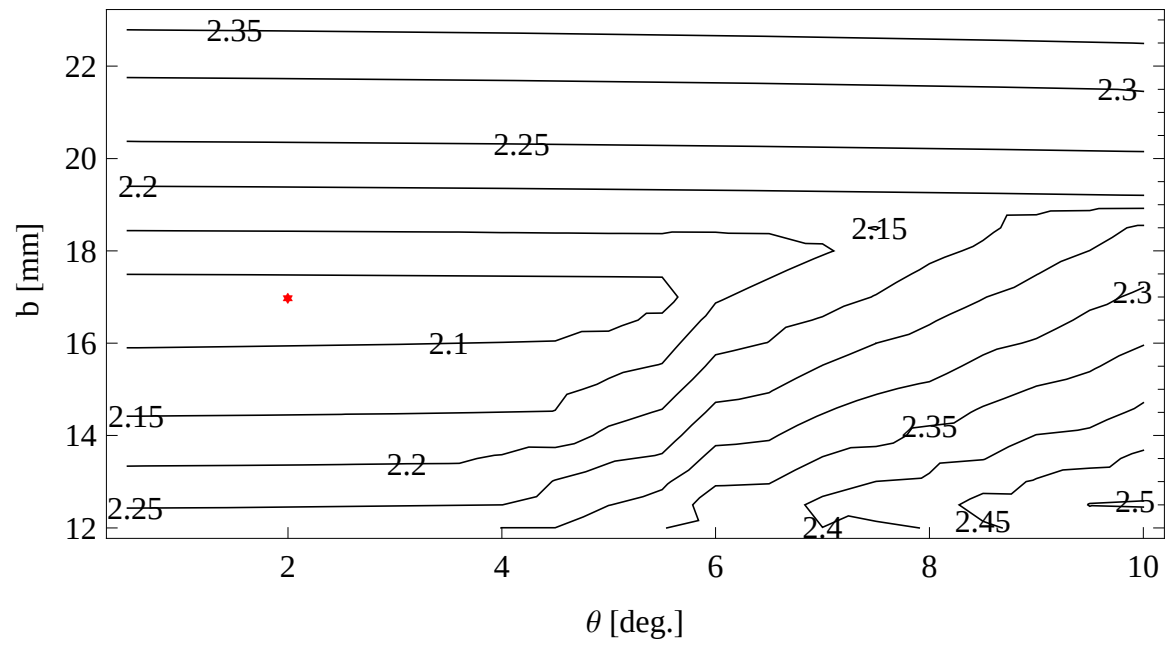

Figure 6: $E_s/E_a$ for various re-entrant shapes. The iris radius and thickness are fixed at 32 mm radius and 21 mm, respectively. The NLSF-RE design is indicated by a single red point.

We also investigated other shapes by varying the angle θ (as indicated in Fig. 1). The characteristic re-entrant-like shape is produced when $\theta < 0$ and the TESLA-like shape $\theta > 0$ (the extant TESLA design is of course the optimum).

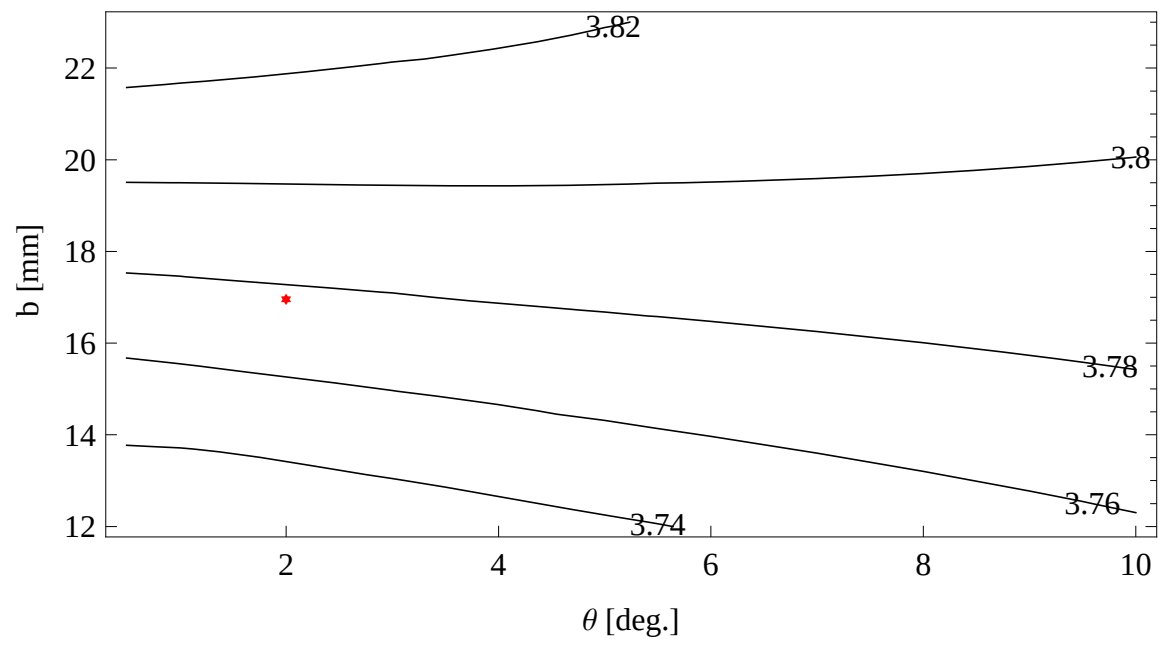

Figure 7: $B_s/E_a$ for various re-entrant shapes with the parameters corresponding to those in Fig 6

Contours of the three figures of merit for the re-entrant-like designs are displayed in Figs 6-8. The optimum design was validated in HFSS and results are the final

Table 1: Comparison of various optimized high gradient cavity designs

| **Parameter** | **TESLA** | **LL** | **ICHIRO** | **NLSF** | **NLSF-A** | **NLSF-RE** |
|---|---|---|---|---|---|---|
| $R_i$ [mm] | 35 | 30 | 30 | 32 | 31 | 32 |
| $R_{eq}$ [mm] | 103.3 | 98.58 | 98.14 | 98.58 | 98.58 | 98.58 |
| A [mm] | 42 | 50.052 | 50.052 | 47.152 | 47.652 | 49 |
| B [mm] | 42 | 36.50 | 34.222 | 31.35 | 32.91 | 35.30 |
| *a* [mm] | 12 | 7.6 | 7.6 | 10.5 | 10 | 10.5 |
| *b* [mm] | 19 | 10 | 9.945 | 15.5 | 15.5 | 17 |
| $f_s$ [GHz] | 1.30116 | 1.30008 | 1.30040 | 1.30023 | 1.30012 | 1.29991 |
| Bandwidth [MHz] | 24.32 | 19.74 | 19.82 | 20.50 | 19.07 | 21.41 |
| $k_c$ [%] | 1.89 | 1.53 | 1.54 | 1.59 | 1.48 | 1.66 |
| $E_s/E_a$ | 2.18 | 2.42 | 2.37 | 2.11 | 2.13 | 2.07 |
| $B_s/E_a$ [mT/(MV/m)] | 4.18 | 3.64 | 3.62 | 3.83 | 3.76 | 3.78 |

optimised cavity, referred to as NLFS-RE, is compared to other designs in Table 1.

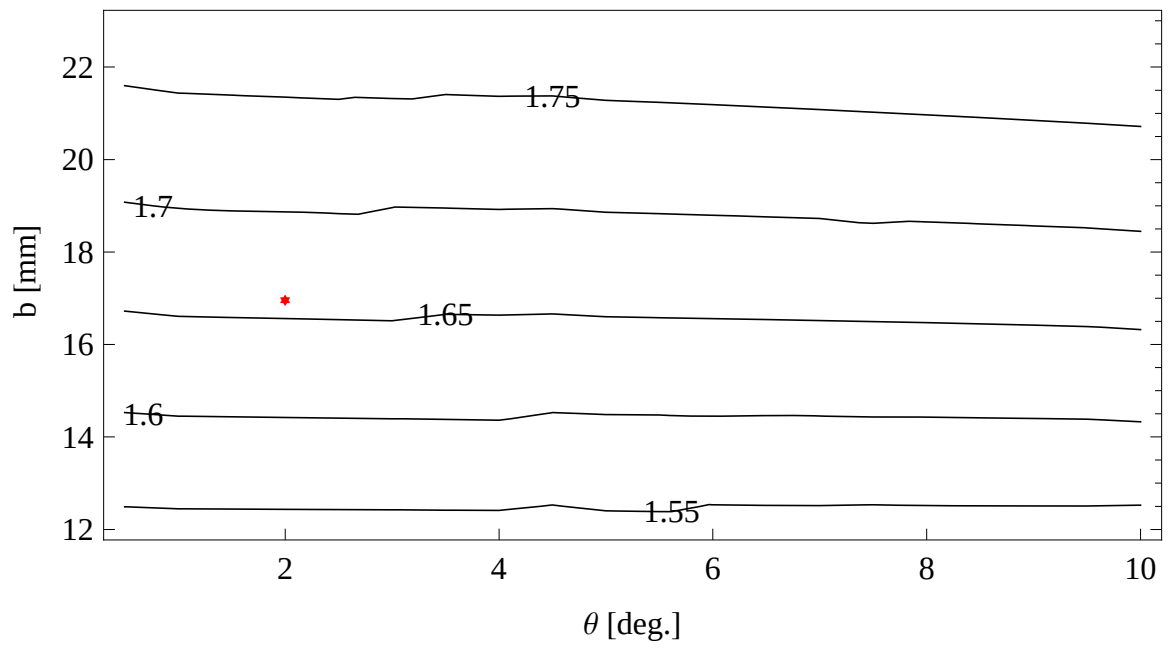

Figure 8: Bandwidth $k_c$ for various re-entrant shapes with the parameters corresponding to those in Fig 6.

The e.m. fields ratios along the surface contour of these optimized three designs are illustrated in Fig. 9, together with the original NLSF [12] design. It is notable that the NLSF middle cell has a 8.5% lower $B_s/E_a$ field compared to TESLA and $E_s/E_a$ is 13% lower than the LL design. Both $E_s/E_a$ and $B_s/E_a$ are comparable to the LSF [11] design. However the bandwidth is superior, as it is ~26.5% wider.

## OVERALL FIELD FLATNESS

We focussed our efforts on a design for the NLSF cavity. A complete design for the NLSF cavity incorporates a study of field flatness and a design of the rf couplers. The radius of the higher order mode (HOM) beam pipes connecting successive cavities to one another, was chosen to be that of the LL design, namely 38 mm, The lowest order mode that can propagate in this beam pipe is a $TE_{11}$ mode, cutoff at ~2.3 GHz. Thus, the first two dipole modes (which are below 2.3 GHz) are contained within the cavity and are damped with suitably designed couplers. The higher order dipole modes can propagate out of the cavity and in this case are partially damped with the HOM couplers.

End cells were carefully designed to ensure a flat accelerating field is obtained. Again, a Mathematica code was written to control Superfish and hence facilitate a rapid variation of parameters. The field flatness is defined as: $\zeta = \left(1 - \sigma_p / \mu_p\right)$. In our simulations we obtained a field flatness of 99%. This result was successfully validated with HFSS simulations.

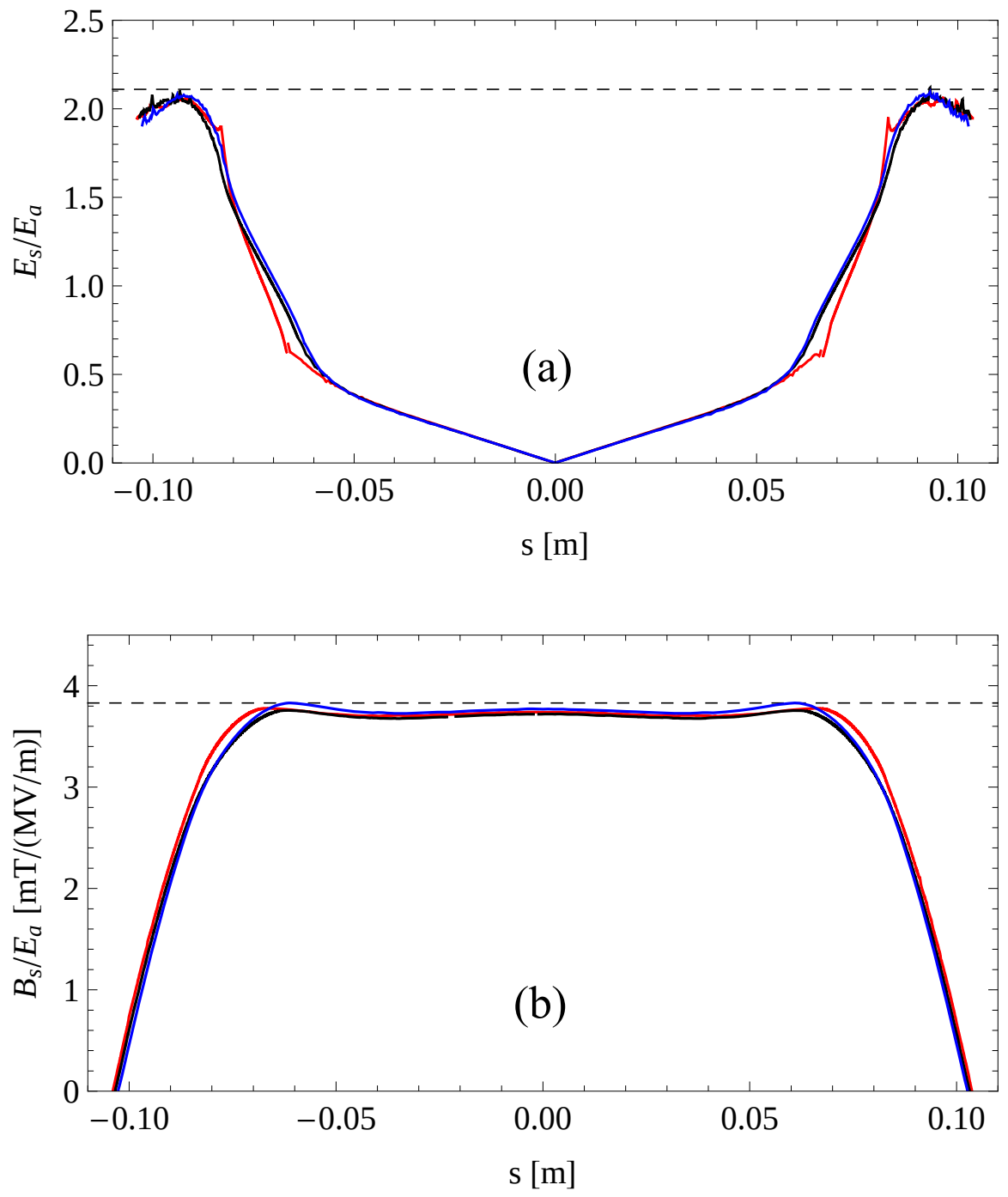

Figure 9: Surface electric field (a) and magnetic field (b) compared to NLSF (blue), NLSF-A (black) and NLSF-RE (red) designs. Here *s* is the coordinate along the surface of the cavity. Dashed lines indicate values of NLSF shape.

The optimized NLSF cavity has a resonant frequency which is within 0.02% of the 1.3 GHz operating frequency and is 99% flat. The field distribution is illustrated in Fig.10. Finally it is to be noted that this optimized end cell is very similar in shape to the middle cells, except for slight difference in the iris and equator elliptical parameters (*a*=9.5, *b*=12.5, *A*=48.152 and *B*=30.5 mm).

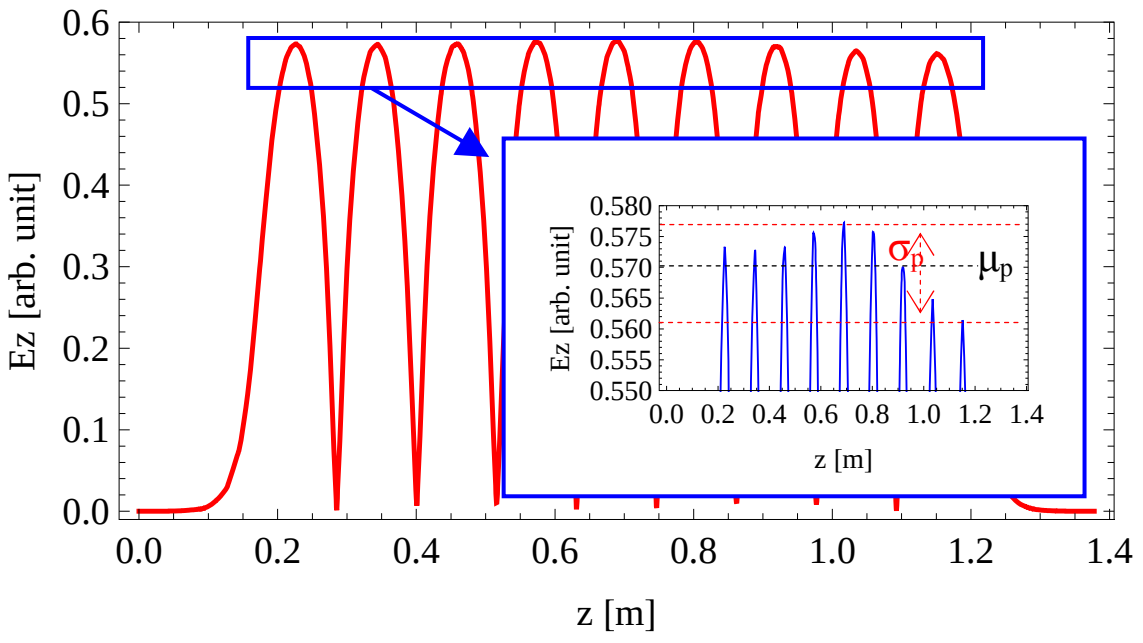

Figure 10: Accelerating field along the 9-cells NLSF cavity axis.

## NLSF CAVITY MODE PROPERTIES

The modal characteristic of a complete nine-cell NLSF cavities have been investigated. To this end we used the eigenmode module of HFSS v11 to obtain the eigenfrequencies and corresponding eigenvectors. In this manner we also were able to obtain the R/Q values. Firstly, we note that the R/Q of the accelerating mode, a measure of efficiency, is slightly larger than that of the TESLA cavity (by ~ 9%). We also studied the dipole and sextupole bands. These modes have the ability to transversely deflect the beam and can at the very least, dilute the emittance of the beam, or can give rise to a damaging beam break up instability. It is important to be able to accurately simulate these modes and ascertain their R/Q values in order to include their effect in beam dynamics simulations.

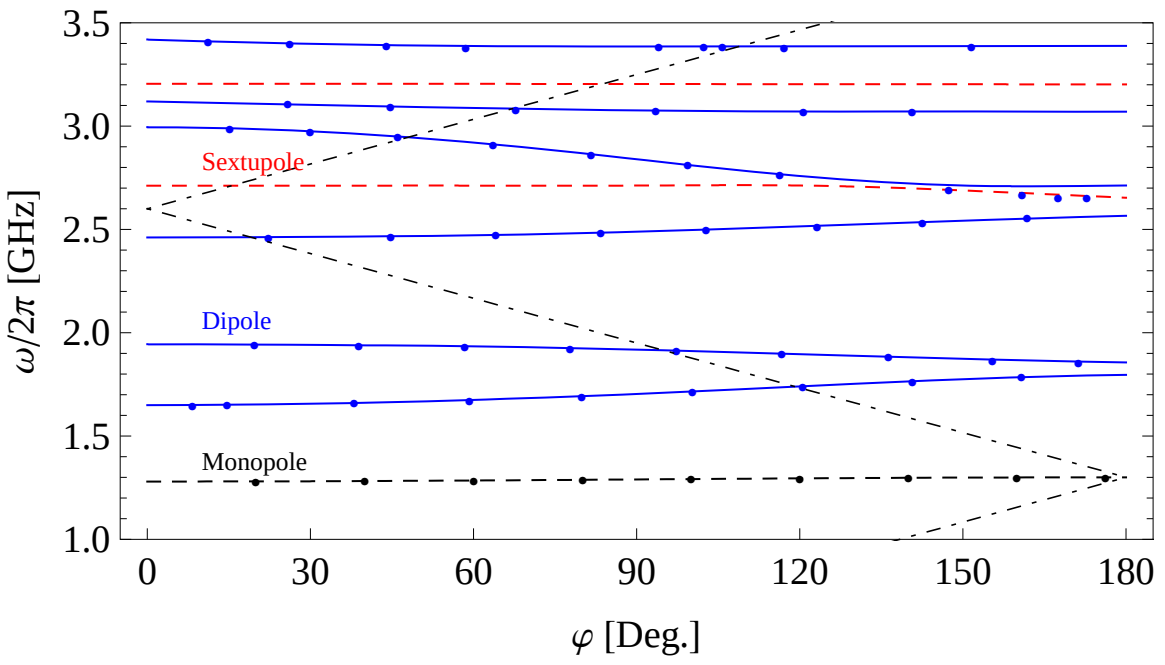

Figure 11: Brillouin diagram of a NLSF cavity. Cells subjected to an infinite periodic structure are indicated by the lines. Monopole, dipole and sextupole modes are indicated by the following lines: black dashed, solid blue and red dashed respectively. The points indicate eigenfrequencies of the complete 9-cell cavity.

A Brillouin diagram of these higher order modes is displayed in Fig 11. We retain modes up to the sixth dipole band. The dipole curves were obtained from simulations entailing a quarter cavity subjected to symmetry conditions appropriate to dipole mode excitation (an electric wall perpendicular to a magnetic wall). The points correspond to the modes in the 9-cell cavity. Provided the frequencies are below the end iris's terminating cut-off frequency, the eigenmodes are relatively independent of the terminating planes at either end of the cavity. In order to assess the impact of these modes on the momentum kick to the beam we also calculated the R/Q factors with a macro written in the post processor of HFSS v11. and these are illustrated for the dipole modes in Fig. 12.

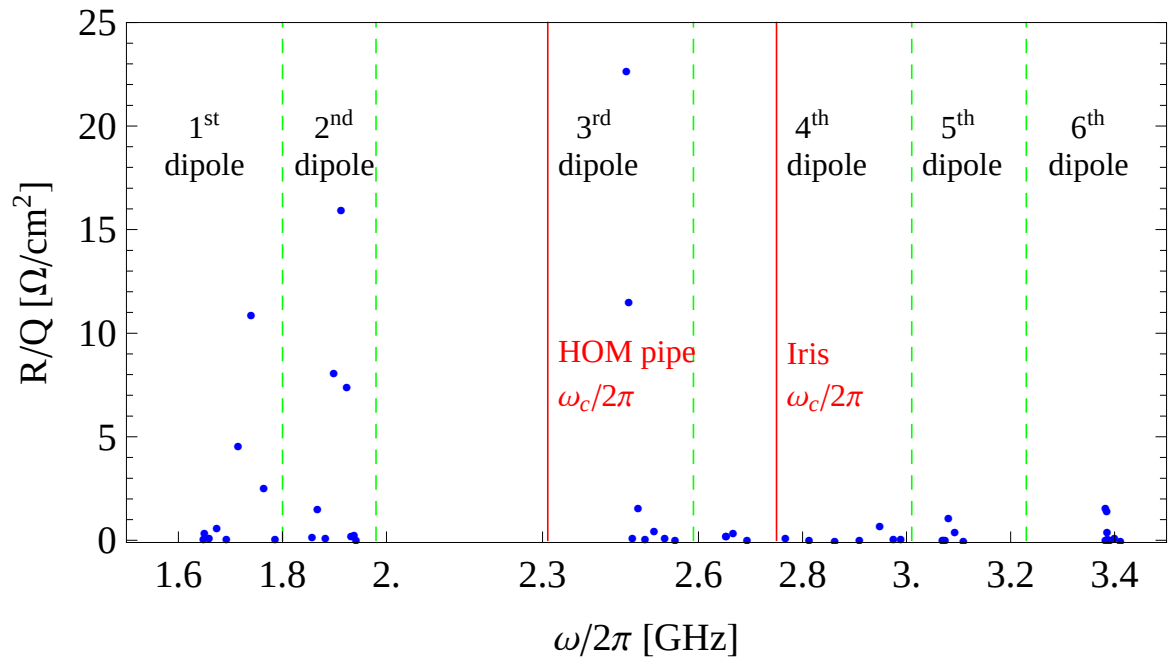

Figure 12: Dipole mode R/Qs for a NLSF cavity including end cells. The cut-off frequencies of the HOM pipe and iris are indicated by $\omega_c/2\pi$.

The modes with large R/Q values are given by those synchronous with the beam and they correspond to the region close to the intersection of the light line with the dipole bands in Fig. 12. For example, the largest R/Qs for each of the first three bands are located at phase advances of $2\pi/3$, $5\pi/9$ and $\pi/9$, with the largest being located in the 3rd dipole band. The LSF structure has a similar mode structure [11]. These results also bear comparison to the simulations made for a 9-cell NLSF cavity made-up of identical cells [12]. The end cells do perturb the overall values, but they remain largely unaffected in the first two dipole bands. However, from the third dipole band onward, the discrepancy in frequencies and R/Qs between the two cavities is significant. The modes with frequencies located in the third band and higher have the ability to propagate out of the confines of the cavity. This gives rise to a strong sensitivity to the boundary conditions used in simulating the eigenmodes. In order to accurately model the modes in these regions radiation boundary conditions should be used. This remains an area for further investigation.

These individual modes affect the long-range wakefield. The short range wakefield, corresponds to a summation of many modes, and is also important as it effects the beam dynamics of along each individual bunches. In this case it is more appropriate to calculate the overall loss factor, rather than individual modal components. We made detailed simulations on the loss factors of our NLSF cavity with ECHO2D [16]. The results of these simulations are summarized in Table 2. For the sake of comparison with existing cavities [9] these simulations were made with a Gaussian bunch of longitudinal length ($\sigma_z$) 1 mm. However, the ILC is designed to be capable of accelerating considerably smaller bunches, of length 300 μm. Additional simulations are included in the table appropriate to ILC bunch lengths.

Table 2: RF parameters of the NLSF cavity.

| **Parameters** | **Unit** | **NLSF** |
|---|---|---|
| Frequency | MHz | 1300 |
| Aperture | mm | 64 |
| Equator | mm | 197.16 |
| HOM pipe aperture | mm | 76 |
| HOM pipe length | mm | 170 |
| $k_c$ | % | 1.59 |
| $B_s/E_a$ | mT/(MV/m) | 2.11 |
| $E_s/E_a$ | - | 3.86 |
| R/Q | Ω | 1127 |
| $k_L$ | V/pC | 11.32 |
| $k_T$ | V/pC/m/m | 22.65 |

## FUNDAMENTAL MODE COUPLER

Preliminary results on simulations of the fundamental mode power coupler, made with the aid of Microwave Studio [17], indicate that a straightforward design, entailing a coaxial type coupler of outer diameter 40 mm and an inner coax of diameter 15 mm is sufficient to achieve an external quality factor ($Q_{ext}$) of $3.5 \times 10^6$. In the MWS simulations this coupler is placed 45 mm away from cavity entrance and the inner coax penetrates to a depth of 6 mm into the cavity. The dependence on external Q factor on proximity to the cavity is displayed in Fig. 15

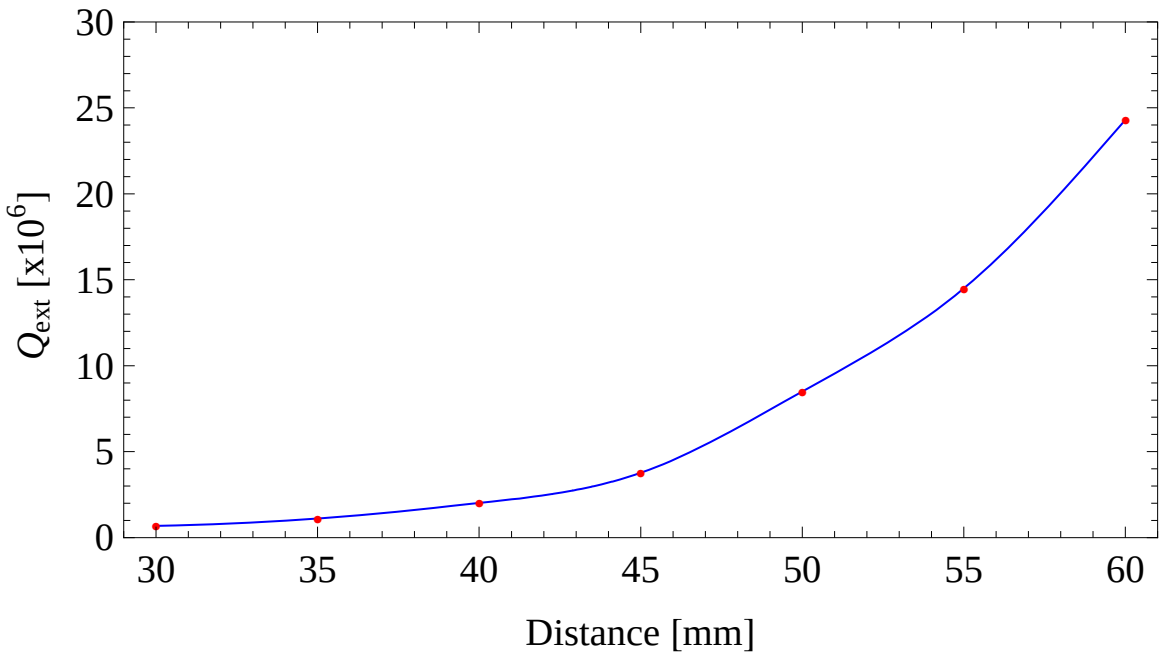

Figure 15: External quality factor versus distance of RF coupler from cavity entrance.

## CONCLUDING REMARKS

The NLSF cavity has the potential to reach accelerating gradients of 60 MV/m and this will give rise to a maximum electric and magnetic field on the surface of 126 MV/m and ~230 mT, respectively. These surface fields, shown in Figs 13-14, are within the acceptable limits set by the critical fields. The accelerating mode R/Q is comparable to the TESLA design (~9% larger). The higher order modes are redistributed but have similar values to the TESLA cavity. Damping these modes is not anticipated to be problematic; although a full study on possible trapped modes is in progress. The external Q of the fundamental mode coupler is satisfactory. A more complete study of both the fundamental and higher order mode coupler for this cavity is in progress. Finally we note that the re-entrant design included also has the potential to achieve low-surface fields.

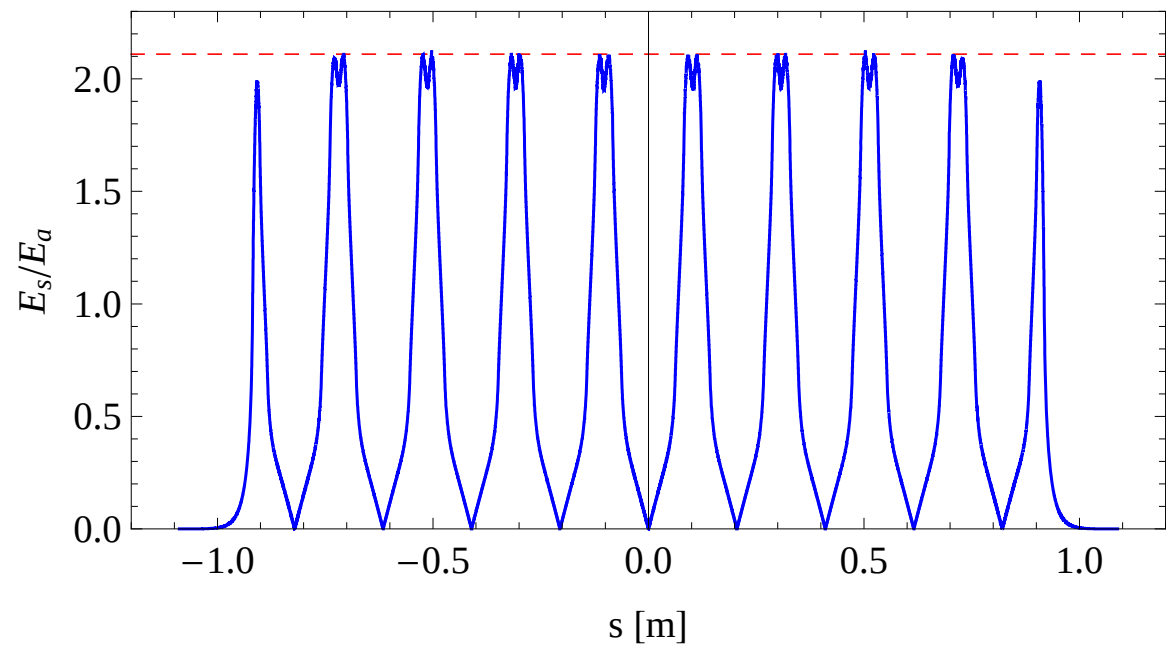

Figure 13: Surface electric field of a NLSF cavity.

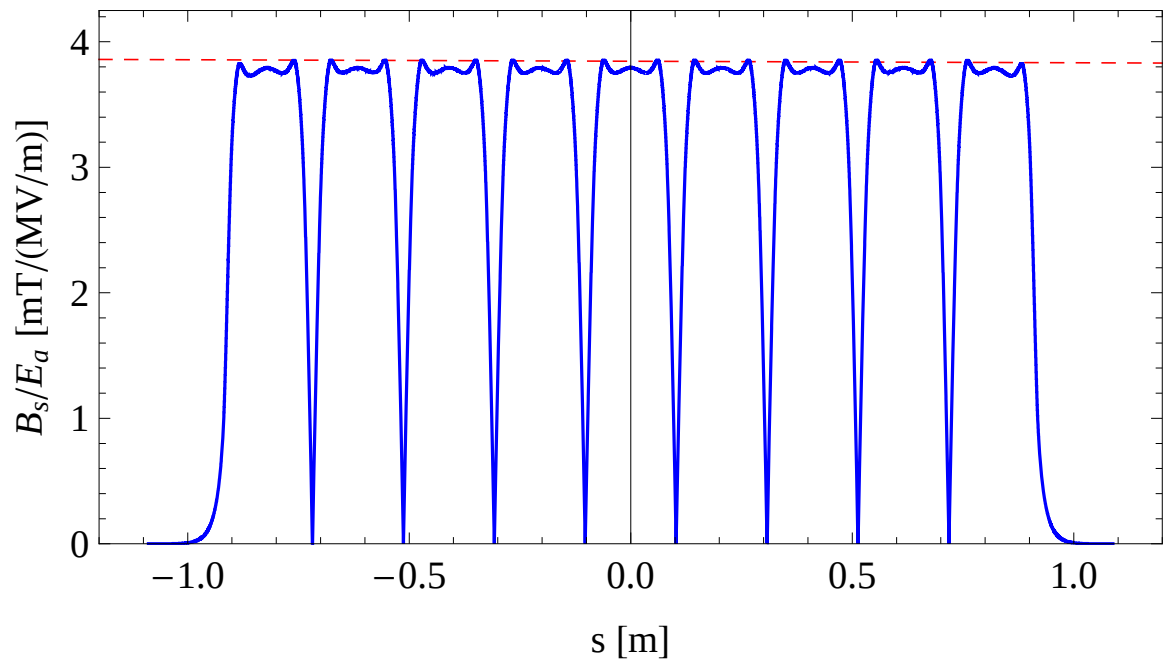

Figure 14: Surface magnetic field of a NLSF cavity.

## ACKNOWLEDGEMENT

We have benefited from discussions at the weekly Manchester Electrodynamics and Wakefield (MEW) meetings held at the Cockcroft Institute, where these results were first presented. N.J. is in receipt of joint support from the Royal Thai Government and the Thai Synchrotron Light Research Institute.